\documentclass[twocolumn,secnumarabic,amssymb,nobibnotes,nofootinbib,aps,prc]{revtex4}

\usepackage{graphicx}
\usepackage{bm}

\newcommand{\half}{{\textstyle{1\over2}}}
\renewcommand{\vec}[1]{\bm{#1}}

\begin{document}

\title{The Roper resonance: a genuine quark state or a dynamically 
generated structure?}

\author{B.~Golli}\email{bojan.golli@ijs.si}
\affiliation{Faculty of Education,
              University of Ljubljana, 1000 Ljubljana, Slovenia}
\affiliation{J.~Stefan Institute,
              1000 Ljubljana, Slovenia}
\author{H. Osmanovi\'c}
\affiliation{Faculty of Natural Sciences and Mathematics, University of Tuzla
              75000 Tuzla, Bosnia and Hercegovina}
\author{S.~\v{S}irca}
\affiliation{Faculty of Mathematics and Physics,
              University of Ljubljana,
              1000 Ljubljana, Slovenia}
\affiliation{J.~Stefan Institute,
              1000 Ljubljana, Slovenia}
\author{A.~\v{S}varc}
\affiliation{Institute Rudjer Bo\v{s}kovi\'c, Zagreb, Croatia}

\date{\today}

\begin{abstract}
In view of the recent results of lattice QCD simulation in the 
P11 partial wave that has found no clear signal for the 
three-quark Roper state we investigate a different mechanism for 
the formation of the Roper resonance in a coupled channel approach 
including the $\pi N$, $\pi\Delta$ and $\sigma N$ channels.
We fix the pion-baryon vertices in the underlying quark model 
while the $s$-wave sigma-baryon interaction is introduced
phenomenologically with the coupling strength, the mass
and the width of the $\sigma$ meson as free parameters.
The Laurent-Pietarinen expansion is used to extract the information 
about the $S$-matrix pole.
The Lippmann-Schwinger equation for the $K$ matrix with a separable
kernel is solved to all orders.
For sufficiently strong $\sigma NN$ coupling the kernel becomes 
singular and a quasi-bound state emerges at around 1.4~GeV, 
dominated by the $\sigma N$ component and reflecting itself in 
a pole of the $S$-matrix.
The alternative mechanism involving a $(1s)^22s$ quark resonant 
state is added to the model and the interplay of the dynamically 
generated state and the three-quark resonant state is studied.
It turns out that for the mass of the three-quark resonant state 
above 1.6~GeV the mass of the resonance is determined solely by 
the dynamically generated state, nonetheless, the inclusion of the 
three-quark resonant state is imperative to reproduce the experimental 
width and the modulus of the resonance pole.
\end{abstract}



\maketitle

\section{Introduction}

Ever since the Roper resonance has been discovered in $\pi N$ 
scattering \cite{roper64}, its exact nature remains unclarified.   
Modern partial-wave analyses \cite{shklyar13,anisovich12a,arndt06} 
of the Roper resonance reveal a non-trivial structure of its poles 
in the complex energy plane, indicating that any kind of Breit-Wigner 
interpretation is inadequate.
The constituent quark model, assuming a (1s)$^2$(2s)$^1$
configuration of the resonance, fails to reproduce many of
the observed properties, in particular in the electro-magnetic sector.  
Various investigations \cite{Capstick,Weber,Cardarelli,Julia-Diaz04,Inna07} 
have emphasized the importance of correct relativistic approach in the 
framework of constituent quark models.  It has also been suggested that 
additional degrees of freedom, such as explicit excitations of the gluon 
field \cite{Li}, the glueball field \cite{broRPA}, or chiral fields 
\cite{Krehl,Roenchen13,Segovia15,Hernandez02,Pedro,EPJ2008,EPJ2009} 
may be relevant for the formation or decay of the Roper resonance.  
The need to include the meson cloud in a quark-model description of the 
Roper resonance 
has been studied in \cite{Cano,Dong,Tiator88,tubingen11,tubingen14}.  
The quark charge densities inducing the nucleon to Roper transition have 
been determined from the phenomenological analysis \cite{Tiator09} 
confirming the existence of a narrow central region and a broad outer band.
In \cite{Krehl} coupled-channel meson-baryon dynamics alone was sufficient 
to engender the resonance; there was no need to include a genuine 
three-quark resonance in order to fit the phase shifts and inelasticities.
This picture has been further elaborated in \cite{Roenchen13}.
Their conclusion may be compared to the EBAC approach 
\cite{Sato-Lee10a,Sato-Lee10b} which emphasizes the important role of the 
bare baryon structure at around 1750~MeV in the formation of the resonance.

The hunt for the Roper is also a perpetual challenge to lattice QCD which 
may ultimately resolve the dilemma about the origin of the resonance.
Although the picture seems to be clearing slowly \cite{leinweber15}, 
the recent calculations of the Graz-Ljubljana group \cite{lang16}
including besides $3q$ interpolating fields also operators for $\pi N$  
in relative $p$-wave and  $\sigma N$ in $s$-wave, and a similar 
calculation by the Adelaide group \cite{kiratidis17} show, however, 
no evidence for a dominant $3q$ configuration below 1.65~GeV and 
2.0~GeV, respectively, that could be interpreted as a Roper state.
The Graz-Ljubljana group has concluded that $\pi N$ channel alone
does not render a low-lying resonance and that coupling with 
$\pi\pi N$ channels seems to be important, which supports the 
dynamical origin of the Roper.
The Adelaide group analysed two scenarios of the resonance formation 
in the framework of the Hamiltonian effective field theory
\cite{liu17,jjwu17}, the dynamical generation and the generation 
through a low-lying bare-baryon resonant state: while both these 
effective models reproduce well the experimental phase shift by 
suitably adjusting model parameters, only the former scenario provides 
an adequate interpretation of the lattice results 
of the Graz-Ljubljana group \cite{lang16}.
The width of the Roper resonance and the quark mass dependence of 
its mass has also been investigated in relativistic baryon ChiPT 
\cite{Gegelia16,Borasoy08}.

In order to investigate the dominant mechanism responsible for 
the resonance formation we devise a simplified model that 
incorporates its dynamical generation as well as the generation
through a three-quark resonant state.
The model is based on our previous calculations covering the
pertinent partial waves and resonances in the low and intermediate 
energy range, and includes only those degrees of freedom which have 
turned out to be the most relevant in this partial wave in the 
energy range of the Roper resonance.
Also, we fix the parameters of the model to the values 
used in our previous calculations, 
except for the $\sigma N$ channel which substantially contributes in
the P11 partial wave and much less in other waves.
This allows us to study the dependence of the results on the 
strength of the $\sigma$ coupling to the nucleon as well as
on the gradual inclusion or exclusion of the three-quark resonant
state.

In the next section we briefly review the basic features 
of our coupled-channel approach and of the underlying quark model.
We construct meson-baryon channel states which incorporate 
the quasi-bound quark-model states corresponding to the nucleon 
and its higher resonances.  
The structure of the multi-channel  $K$ matrix is discussed and 
the method to solve the Lippmann-Schwinger equation for the meson 
amplitudes is outlined.
In Sect.~\ref{sec:withoutR} we solve the coupled-channel problem
without including the genuine Roper state.   In Sect.~\ref{sec:withR} 
the problem is solved by explicitly including the corresponding 
three-quark resonant state.

\section{\label{model} The model}

\subsection{\label{qmodel} The underlying quark model}

In our approach to scattering and photoproduction of mesons
we use a chiral quark model to determine the meson-baryon and 
photon-baryon vertices, which results in a substantially smaller 
number of free parameters compared to more phenomenological 
methods used in the partial-wave analyses.
We use a $SU(3)$ extended version of the Cloudy Bag Model (CBM) 
\cite{CBM} supplemented with addition of the $\sigma$, $\rho$ 
and $\omega$ mesons.
The bag radius $R=0.83$~fm corresponding to the cut-off 
$\Lambda\approx 550$~MeV and value of the pion-decay constant, 
$f_\pi$, reduced to 76~MeV in order to reproduce the $\pi NN$ 
coupling constant and other nucleon ground-state properties 
have been used for all pertinent resonances.
In this model we have analysed the $S$, $P$ and $D$ partial waves
including all relevant resonances and channels in the low- and 
intermediate-energy regime 
\cite{EPJ2008,EPJ2009,EPJ2011,EPJ2013,EPJ2016}.
For most of the $S$ and $P$ resonances the parameters determined in 
the underlying quark model describe consistently the scattering 
and photo-production amplitudes, including the production of 
$\eta$ and $K$ mesons.
The model, however, underestimates the $d$-wave meson coupling to 
the quark core, typically by a factor of two.

In the present calculation we have included  the channels
that in our previous calculations turned out to be the most 
relevant in the energy range of the Roper resonance: apart from
the elastic channel, the $\pi\Delta$ and the $\sigma N$ channels.
From our experience mostly in the P33 wave we have been able
to fix the pion-baryon vertices while the $s$-wave  $\sigma NN$,  
$\sigma\Delta\Delta$ and  $\sigma NR$ vertices, mimicking the 
$\pi\pi$-baryon interactions, are introduced phenomenologically.
We assume
$$
   V^\sigma_0(k,\mu) 
   = V^\sigma_0(k)\,w_\sigma(\mu)    \,,
\quad
    V^\sigma_0(k) 
   = g\,{k\over\sqrt{2\omega(k)}}\,.
$$
Here $\omega^2(k) = k^2 + \mu^2$, $\mu$ is the invariant mass of the 
two-pion system and $w_\sigma(\mu)$ is a Breit-Wigner function 
centered around $\mu_\sigma$ with the width $\Gamma_\sigma$; 
$g$ as well as the Breit-Wigner values are free parameters of the model
and are assumed to be the same for all three $\sigma$ vertices.
We use $g$ to study the behaviour of the amplitudes in different regimes.
Our calculation favours  $\mu_\sigma$ and $\Gamma_\sigma$ which are 
slightly larger than the PDG values of $\mu_\sigma\approx 450$~MeV and 
$\Gamma_\sigma\approx 500$~MeV \cite{PDG}.
We present results for two pairs of values,
$\mu_\sigma=\Gamma_\sigma= 500$~MeV and $\mu_\sigma=\Gamma_\sigma= 600$~MeV.
For the background we include only the $u$-channel processes
with the intermediate nucleon and $\Delta(1232)$ which, based on our 
previous experience in the P11 and also in the P33 partial wave,
dominates in the considered energy regime.

\subsection{The coupled channel approach}

In our approach the quasi-bound quark state is included through 
a scattering state which in channel $\alpha$ assumes the form
\begin{eqnarray}
   |\Psi_\alpha\rangle &=& \mathcal{N}_\alpha\biggl\{
        \left[a^\dagger_\alpha(k_\alpha)|\Phi_\alpha\rangle\right]
   + c_{\alpha N}|\Phi_N\rangle +  c_{\alpha R}|\Phi_R^0\rangle
\biggr.\nonumber\\ && \biggl. 
  +  \sum_\beta
   \int 
       {\mathrm{d}k\>\chi_{\alpha\beta}(k_\alpha,k)\over\omega_\beta(k)+E_\beta(k)-W}\,
      \left[a^\dagger_\beta(k)|\Phi_\beta\rangle\right]\biggr\},
\label{PsiH}
\end{eqnarray}
where $\alpha$ ($\beta$) denotes either $\pi N$, $\pi\Delta$ or $\sigma N$ 
channels, and [ ] stands for coupling to total spin $\half$ and isospin $\half$.
The first term represents the free meson ($\pi$ or $\sigma$) and the baryon 
($N$ or $\Delta$) and defines the channel, the next term is the admixture 
of the nucleon ground state, the third term corresponds to a {\em bare\/} 
three-quark state, while the fourth term describes the pion cloud around 
the nucleon and $\Delta$, and the $\sigma$ cloud around the nucleon.
Here $\mathcal{N}_\alpha=\sqrt{\omega_\alpha E_\alpha/(k_\alpha W)}$, $k_\alpha$ 
and $\omega_\alpha$ are on-shell values,\footnote{%
In the following we use $\mu_\alpha$ for the meson mass in channel $\alpha$, 
such that $\omega_\alpha^2(k)=k^2+\mu_\alpha^2$; the vertex in an $u$-channel 
exchange is denoted by $V^\beta_{i\alpha}$, with $\beta$ corresponding to the 
meson in channel $\beta$, the vertex in a $s$-channel by $V_{\alpha B}$.}
where  $W=\omega_\alpha+E_\alpha$ is the invariant mass.

The (half-on-shell) $K$ matrix is related to the scattering state 
as \cite{EPJ2008}
\begin{equation}
   K_{\alpha\beta}(k_\alpha,k) = -\pi\mathcal{N}_\beta
  \langle \Psi_\alpha||V^\beta(k)||\Phi_\beta\rangle\,,
\end{equation}   
with the property $K_{\alpha\beta}(k_\alpha,k) = K_{\beta\alpha}(k,k_\alpha)$.
It is proportional to the meson amplitudes $\chi$ in (\ref{PsiH}),
\begin{equation}
   K_{\alpha\beta}(k_\alpha,k)  
       = \pi\,\mathcal{N}_\alpha\mathcal{N}_\beta\,
             \chi_{\alpha\beta}(k_\alpha,k) \,.
\label{chi2K}
\end{equation}
The equations for the meson amplitudes can be derived from requiring the 
stationarity of the functional, $\langle\delta\Psi|H-W|\Psi\rangle=0$, 
which leads to the Lippmann-Schwinger type of equation
\begin{eqnarray}
   \chi_{\alpha\gamma}(k,k_\gamma) 
   &=& -{c}_{\gamma N}\, {V}_{\alpha N}(k)-{c}_{\gamma R}\, V_{\alpha R}(k)
       + \mathcal{K}_{\alpha\gamma}(k,k_\gamma)
\nonumber\\ && 
+ \sum_\beta\int\mathrm{d}k'\;
  {\mathcal{K}_{\alpha\beta}(k,k')\chi_{\beta\gamma}(k',k_\gamma)
  \over \omega_\beta(k') + E_{\beta}(k')-W}\,.
\label{eq4chi}
\end{eqnarray}
The kernel (averaged over meson directions) reads \cite{EPJ2008}
\begin{equation}
  \mathcal{K}_{\alpha\beta}(k,k') =  
  \sum_{i=N,\Delta} f^i_{\alpha\beta}\, 
  {V_{i\beta}^{\alpha}(k)\,V_{i\alpha}^{\beta}(k')
   \over \omega_\alpha(k)+\omega_\beta(k')+E_i(\bar{k})-W}\,,
\label{kernel}
\end{equation}
where for channels involving pions the spin-isospin factors equal 
$$
   f^N_{NN} = f^\Delta_{NN}={1\over9},\; 
   f^\Delta_{NN} = f^\Delta_{\Delta\Delta} = {4\over9},
$$
$$
    f^N_{N\Delta} = f^N_{\Delta N}={8\over9},\;
    f^\Delta_{N\Delta} = f^\Delta_{\Delta N}={5\over9}\,,
$$
and 1 if at least one of the channels is $\sigma N$.
As discussed in our earlier work, Eq. (\ref{kernel}) implies dressed 
vertices; in the present calculation the vertices involving the 
$\Delta$ are increased by 35~\% to 40~\%  with respect to their bare 
values in accordance with our analysis of the P33 resonances in 
\cite{EPJ2008}, while $V_{\pi NN}$ is kept at its bare value.
We assume the following factorization of the denominator in 
(\ref{kernel}):
\begin{eqnarray}
&&  \kern-30pt {1 \over \omega_\alpha(k)+\omega_\beta(k')+E_i-W} 
\approx    
\nonumber \\ &&
   {(\omega_\alpha + \omega_\beta + E_i - W)
\over 
   (\omega_\alpha(k)+E_i-E_\beta)(\omega_\beta(k')+E_i-E_\alpha)}\,,
\label{separable}
\end{eqnarray}
where $W = E_\alpha + \omega_\alpha = E_\beta + \omega_\beta$.
The factorization is exact if either of the $\omega$'s is
on-shell, i.e. $\omega_\alpha(k)\to\omega_\alpha=W-E_\alpha$ or 
$\omega_\beta(k')\to\omega_\beta=W-E_\beta$.
Assuming (\ref{separable}) the kernel can be written in the form
\begin{eqnarray}
   \mathcal{K}_{\alpha\beta}(k,k') &=& \sum_i\varphi^{\alpha}_{\beta i}(k)\,
                                      \xi^{\beta}_{\alpha i}(k')\,,
\label{Ksep}\\
   \varphi^{\alpha}_{\beta i}(k) &=& 
   {m_i\over E_\beta}\,
   (\omega_\beta + \varepsilon^\beta_{i\alpha})
   {V^\alpha_{i\beta}(k)\over\omega_\alpha(k) +\varepsilon_{i\beta}^\alpha}\;
    f^i_{\alpha\beta}\,,
\nonumber\\
   \xi^{\beta}_{\alpha i}(k') &=&
   {V^\beta_{i\alpha}(k')\over\omega_\beta(k') +\varepsilon_{i\alpha}^\beta}\,.
\nonumber
\end{eqnarray}
We further modify the propagator $(\omega_\beta(k')+E_i-E_\alpha)^{-1}$
such that it corresponds to the denominator in the $u$ channel,
i.e. $2m_i/(2E_\alpha\omega_\beta(k)+m_i^2-m_\alpha^2 -\mu_\beta^2)$
in which case
\begin{equation}
  \varepsilon^\beta_{i\alpha} = 
        {m_i^2 - m_\alpha^2 - \mu_\beta^2 \over 2E_\alpha}\,,
\end{equation}
with the property 
$E_\alpha( \omega_\beta  + \varepsilon^\beta_{i\alpha})
 = E_\beta(\omega_\alpha + \varepsilon^\alpha_{i\beta})$.
Let us note that the half-on-shell $\mathcal{K}_{\alpha\beta}(k,k_\beta)$
reduces to the standard form of the $u$-channel background in
the so-called Born approximation:
\begin{equation}
   \mathcal{K}_{\alpha\beta}(k,k_\beta) = \sum_if^i_{\alpha\beta}\,
   {2m_i\,V^\alpha_{i\beta}(k)V^\beta_{i\alpha}(k_\beta)\over
    2E_\alpha\omega_\beta +  m_i^2 - m_\alpha^2 - \mu_\beta^2}\,.
\label{Khalfon}
\end{equation}
The above approximations are discussed in more detail in 
Appendix C of \cite{EPJ2008}.

\section{\label{sec:withoutR} Solution without three-quark resonant states}

\begin{figure}[h]
$$\includegraphics[width=88mm]{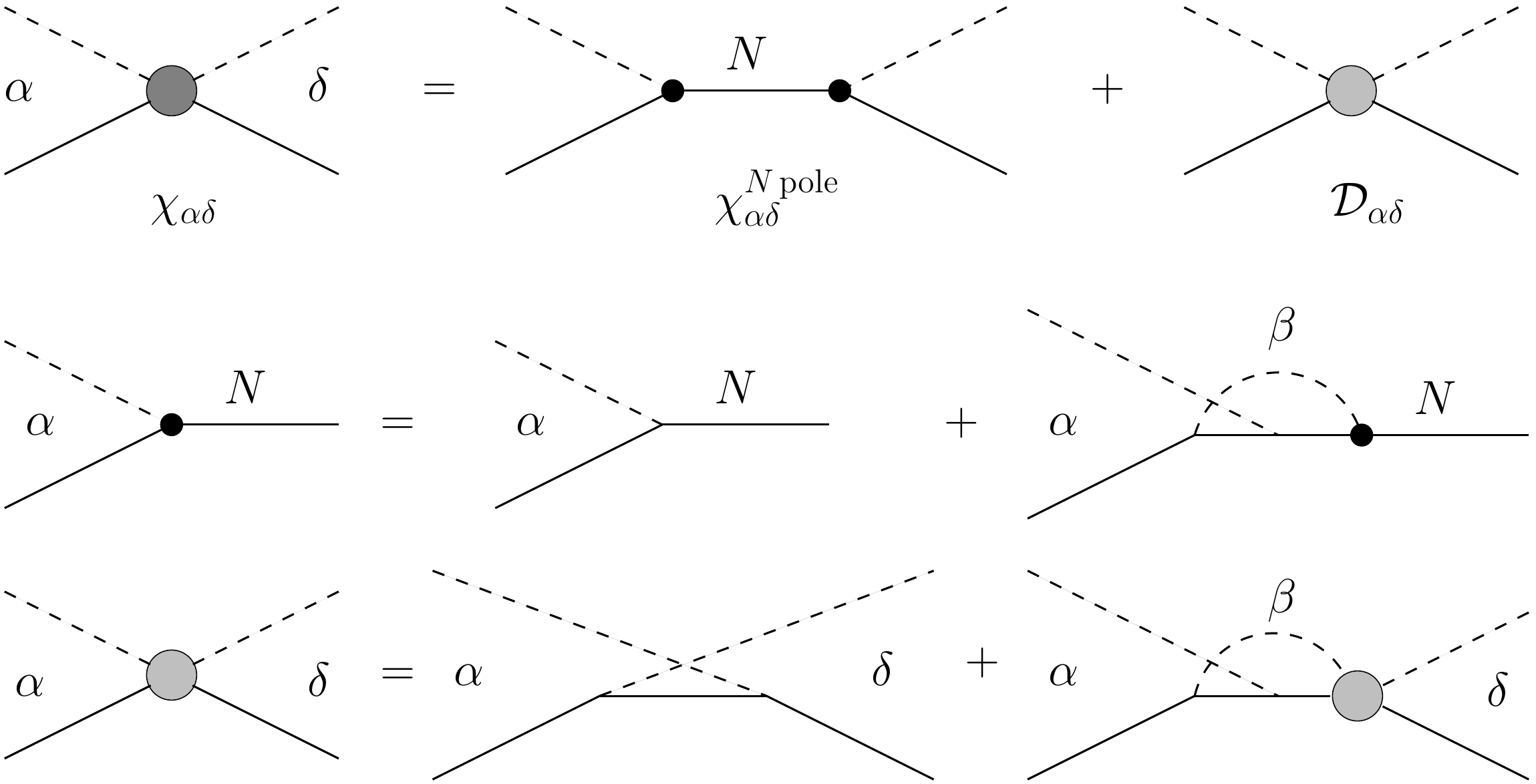}$$
\vspace{-12pt}
\caption{Graphical representation of Eqs. (\ref{splitchi}),
(\ref{eq4VN}) and (\ref{eq4D}) (from top to bottom).} 
\label{fig:LSE}
\end{figure}

The meson amplitude (or equivalently the $K$ matrix) consists of the 
nucleon-pole term and the background (see Fig.~\ref{fig:LSE}):
\begin{equation}
   \chi_{\alpha\delta}(k,k_\delta) = 
    c_{\delta N}{\cal V}_{\alpha N}(k)+
    {\cal D}_{\alpha\delta}(k,k_\delta)\,.
\label{splitchi}
\end{equation}
Equation (\ref{eq4chi}) can be split into the equation for the
dressed vertex,
\begin{equation}
\mathcal{V}_{\alpha N}(k)
= \bar{V}_{\alpha N}(k)
 + \sum_{\beta}  \int\mathrm{d}k'\;
        {\mathcal{K}_{\alpha\beta}(k,k')
         \mathcal{V}_{\beta N}(k')
   \over   \omega_\beta(k')+E_\beta(k')-W}\,,
\label{eq4VN}
\end{equation}
and the background:
\begin{equation}
\mathcal{D}_{\alpha\delta}(k,k_\delta) = 
\mathcal{K}_{\alpha\delta}(k,k_\delta)
+  \sum_\beta \int\mathrm{d}k'\;
        {\mathcal{K}_{\alpha\beta}(k,k')
         \mathcal{D}_{\beta\delta}(k',k_\delta)
   \over   \omega_\beta(k')+E_\beta(k')-W}\,.
\label{eq4D}
\end{equation}
Since we assume that the nucleon in (\ref{PsiH}) is an exact solution of 
the Hamiltonian and therefore consists of the  pion and $\sigma$ clouds, 
the vertex $\bar{V}_{\alpha N}$ in (\ref{eq4VN}) is a modification of the 
ground-states meson amplitudes such that its off-shell value goes to zero 
as $W$ approaches $m_N$:
\begin{equation}
 \bar{V}_{\alpha N}(k) =  
     {(W-m_N)V_{\alpha N}(k)\over\omega_\alpha(k)+E_\alpha(k) - m_N}\,.
\label{barVN}
\end{equation}
Finally
\begin{equation}
    c_{\alpha N} = -{\mathcal{V}_{\alpha N}(k)\over\lambda_N(W)}\,,
\label{eq4cN}
\end{equation}
where
\begin{equation}
   \lambda_N(W) = W - m_N 
       + \sum_\beta\int\mathrm{d}k\;{\mathcal{V}_{\beta N}(k)\bar{V}_{\beta N}(k)
                \over \omega_\beta(k)+E_\beta(k)-W}\,.
\label{eq4lamN}
\end{equation}
Note that due to (\ref{barVN}) the self-energy term vanishes 
as $W\to m_N$.

Since the kernel (\ref{Ksep}) is separable, Eqs. (\ref{eq4VN}) 
and (\ref{eq4D})  can be exactly solved with the ansaetze
\begin{equation}
\mathcal{V}_{\alpha N}(k) = \bar{V}_{\alpha N}(k) + 
  \sum_{\beta i} x^\alpha_{\beta i}\,\varphi^\alpha_{\beta i}(k)
\end{equation}
and
\begin{equation}
\mathcal{D}_{\alpha\delta}(k,k_\delta) =
\mathcal{K}_{\alpha\delta}(k,k_\delta) 
+ \sum_{\beta i} z^{\alpha\delta}_{\beta i}\, \varphi^\alpha_{\beta i}(k)\,.
\end{equation}
This leads to a set of linear algebraic equations for the coefficients 
$x$ and $z$:
\begin{eqnarray}
x^{\alpha}_{\beta i} 
   + \sum_{\gamma j} M^\beta_{\alpha i,\gamma j}\;x^\beta_{\gamma j}
&=& b^\beta_{\alpha i}\,, 
\label{eq4x}\\
 z^{\alpha\delta}_{\beta i}
+ \sum_{\gamma j} M^\beta_{\alpha i,\gamma j}\; z^{\beta\delta}_{\gamma j}  
&=&
 d^{\beta\delta}_{\alpha i}\,, 
\label{eq4z}
\end{eqnarray}
where
\begin{eqnarray}
M^\beta_{\alpha i,\gamma j} 
&=& -\int\mathrm{d}k\;{\xi^{\beta}_{\alpha i}(k)\varphi^{\beta}_{\gamma j}(k)
    \over \omega_\beta(k)+E_\beta(k)-W}\,,
\label{eq4M}\\
b^\beta_{\alpha i}
&=& \int\mathrm{d}k\;{\bar{V}_{\beta N}(k)\xi^{\beta}_{\alpha i}(k)
    \over  \omega_\beta(k)+E_\beta(k)-W}\,,
\nonumber\\
  d^{\beta\delta}_{\alpha i} &=&
\int\mathrm{d}k\;{\mathcal{K}_{\beta\delta}(k,k_\delta)\xi^{\beta}_{\alpha i}(k)
    \over  \omega_\beta(k)+E_\beta(k)-W}\,.
\nonumber
\end{eqnarray}

\begin{figure}[h]
$$\includegraphics[width=60mm]{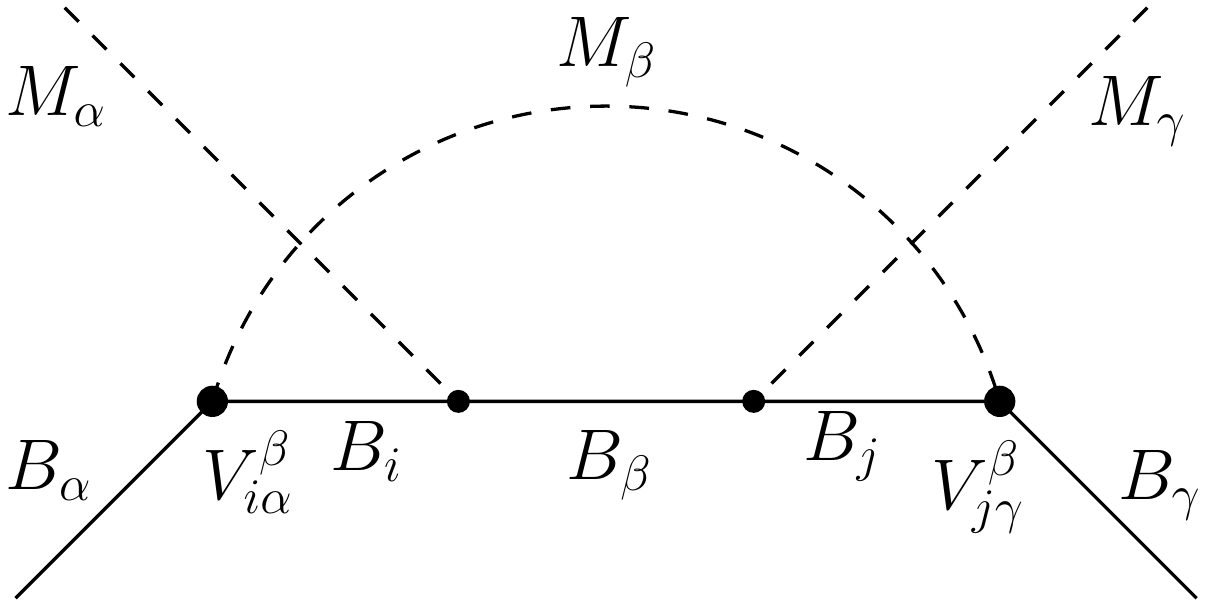}$$
\vspace{-12pt}
\caption{Graphical representation of $M^\beta_{\alpha i,\gamma j}$:
$M_\alpha$ and $B_\alpha$ denote respectively the meson ($\pi$ 
or $\sigma$) and baryon ($N$ or $\Delta$) in channel $\alpha$.}
\label{fig:Agraph}
\end{figure}
The meaning of the indices in the matrix 
${\sf M}=[M]^\beta_{\alpha i,\gamma j}$ is explained in Fig.~\ref{fig:Agraph}.
Introducing ${\sf A} = {\sf I} + {\sf M}$, (\ref{eq4x}) can be written 
as ${\sf A}\vec{x} = \vec{b}$ or, in terms of components, as
\begin{equation}
  \sum_{\gamma j} A^{\beta\eta}_{\alpha i,\gamma j}\;x^\eta_{\gamma j} 
               = b^\beta_{\alpha i}\,,
\end{equation}
where
$$
  A^{\beta\eta}_{\alpha i,\gamma j} = 
   I^{\beta\eta}_{\alpha i,\gamma j} 
+ \delta_{\beta,\eta}M^\beta_{\alpha i,\gamma j}\,,
\qquad
  I^{\beta\eta}_{\alpha i,\gamma j} 
  = \delta_{\alpha,\eta}\delta_{\beta,\gamma}\delta_{i,j}\,.
$$
The set (\ref{eq4z}) for the $z$ parameters differs only in the
right-hand side.
The structure of {\sf A} for our choice of channels is
displayed in the Appendix.

In order to analyze the behaviour of the kernel  we perform
the singular value decomposition of {\sf A},
\begin{equation}
  {\sf A} = {\sf U} {\sf W} {\sf V}^T \,,
\label{svdcmp}
\end{equation}
where {\sf U} and {\sf V} are orthogonal matrices and {\sf W}
is a diagonal matrix.
Since ${\sf A}^{-1} = {\sf V} {\sf W}^{-1} {\sf U}^T$,
the solution for $x$  can be written as
\begin{equation}
   \vec{x} = {\sf A}^{-1}\vec{b}\,,
\qquad
    x_p = \sum_{r=1}^{\mathrm{dim}({\sf A})} {1\over w_r}
            \left(\sum_qU_{rq}b_{q}\right)
                    V_{pr}\,.
\label{sol4x}
\end{equation}
We have introduced common indices $p,q,r$ numbering possible
combinations of the two channels indices (e.g. $\alpha$ and $\beta$) 
and the index of the intermediate state ($i$); in the present
case $\mathrm{dim}({\sf A})=13$.
\begin{figure}[h]
\begin{center}
\includegraphics[width=80mm]{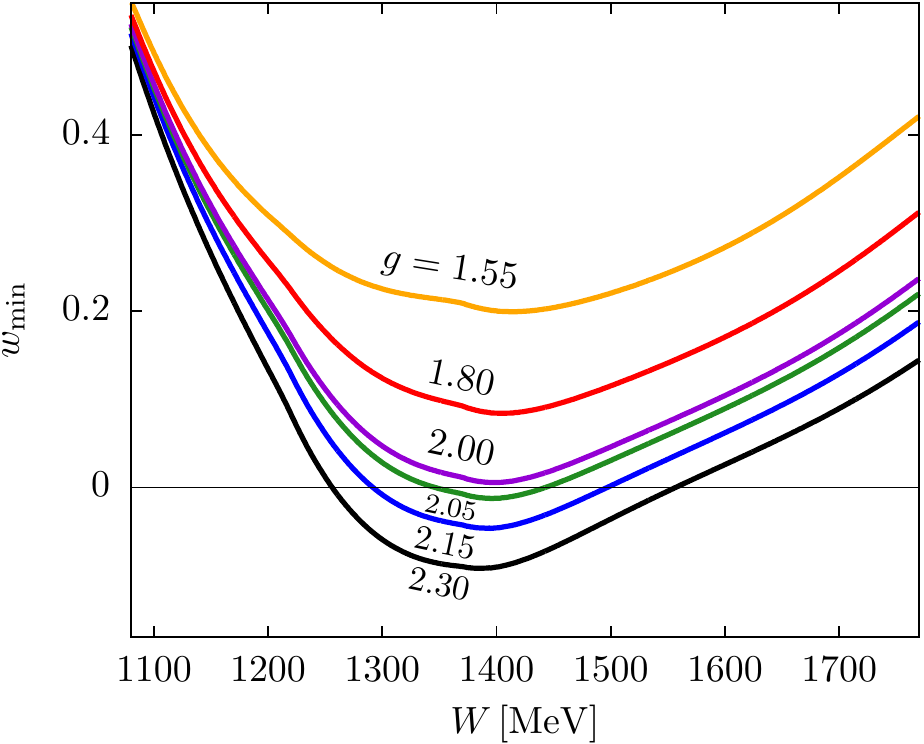}
\end{center}
\vspace{-12pt}
\caption{The behaviour of the lowest eigenvalue $w_{\mathrm{min}}$ as a 
function of $W$ for $\mu_\sigma=\Gamma_\sigma=600$~MeV 
for coupling constant $g$ from 1.55 to 2.3.}
\label{fig:wmin}
\end{figure}

\begin{figure}[h]
\begin{center}
\includegraphics[width=80mm]{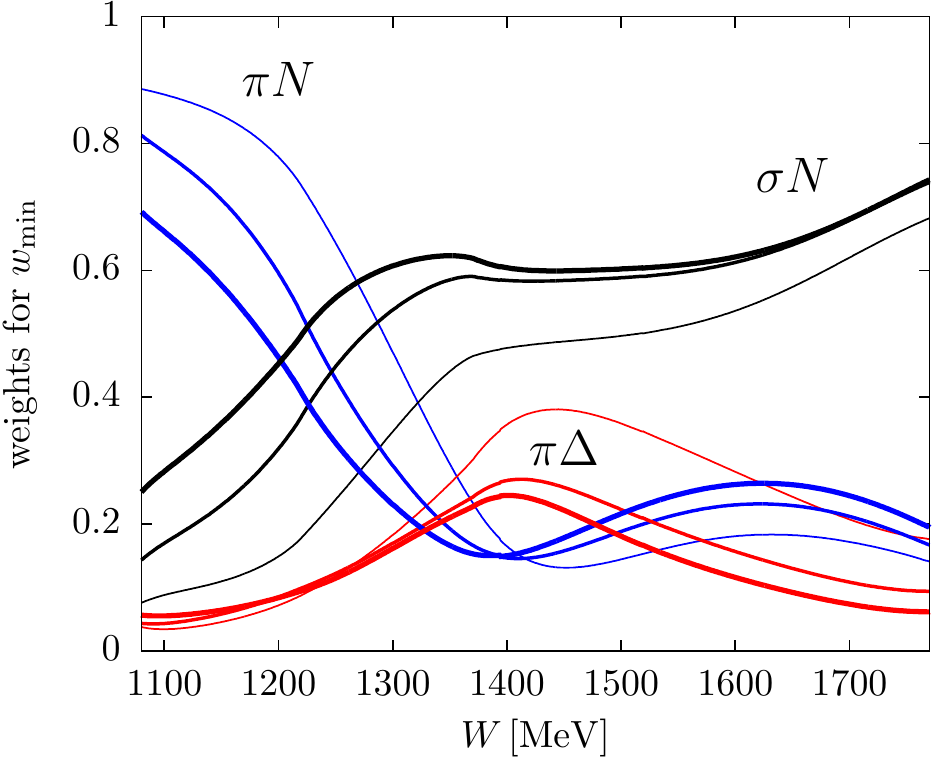}
\end{center}
\vspace{-12pt}
\caption{The  weights of the $\pi N$, $\pi\Delta$ and $\sigma N$
components of the eigenvector belonging to the lowest eigenvalue  
$w_{\mathrm{min}}$ for $g=2.3$ (thick lines), $g=2.0$ (medium lines) 
and $g=1.55$ (thin lines), and $\mu_\sigma=\Gamma_\sigma=600$~MeV.}
\label{fig:vmin}
\end{figure}

For weak couplings $g$ the eigenvalues%
\footnote{~Strictly speaking, the singular-value decomposition
of ${\sf A}$ produces singular values $w_r$ which are square 
roots of the eigenvalues of ${\sf A}{\sf A}^T$.  For simplicity,
we call $w_r$ eigenvalues and the corresponding columns of ${\sf U}$
and ${\sf V}$ eigenvectors.}
$w_r$ remain close to 1;
increasing $g$, one of the eigenvalues denoted $w_{\mathrm{min}}$ becomes 
smaller while the others stay close to 1.
Figure~\ref{fig:wmin} shows that the minimum is reached around 
$W=1400$~MeV almost independently of $g$; this value of the invariant 
mass is also insensitive to the choice of the Breit-Wigner mass and 
width of the $\sigma$ meson.
The solution in the energy range from $W\approx 1300$~MeV to 1600~MeV
is therefore dominated by the lowest eigenvector of {\sf A}.
Around $g=2$, $w_{\mathrm{min}}$  touches zero and beyond this (critical) 
value it crosses zero twice, at $W_1$ and $W_2$.
At these two energies, a nontrivial solution of the homogeneous
equation appears, signaling the emergence of a (quasi)-bound state.
The corresponding eigenvector changes little with $W$ and stays almost 
constant for $1300~\mathrm{MeV}<W<1600~\mathrm{MeV}$.
This remains true even for $g$ smaller than the critical value,
except that the $\sigma N$ component is less strong.
(See Fig.~\ref{fig:vmin}.)
As we shall justify in the following we can identify the corresponding
state as the {\em dynamically generated\/} state.

Let us consider how the lowest eigenvalue influences the behaviour
of the $K$ matrix, which has the form
\begin{equation}
  K_{\alpha\beta}
  = \pi\,\mathcal{N}_\alpha\mathcal{N}_\beta\left[
    {{\cal V}_{\alpha N}{\cal V}_{\beta N}\over \lambda_N(W)}
   + {\cal D}_{\alpha\beta}\right]\,.
\label{KnoR}
\end{equation}
When $w_{\mathrm{min}}$ approaches zero, ${\cal V}_{\alpha N}$ and 
${\cal D}_{\alpha\beta}$ are dominated by the dynamically generated state
and both quantities become proportional to $w_{\mathrm{min}}^{-1}$;
the same is true also for $\lambda_N(W)$ (see Eq.~(\ref{eq4lamN})).
As a result, the $K$ matrix also behaves like $w_{\mathrm{min}}^{-1}$ 
and therefore possesses the pole at those $W$ where $w_{\mathrm{min}}$ 
crosses zero.
Let us note that while the {\sf A} is not symmetric, the resulting
$K$ matrix (\ref{KnoR}) is symmetric (and real), which guarantees 
the unitarity of the $S$ matrix.

\begin{figure}[h]
\begin{center}
\includegraphics[width=80mm]{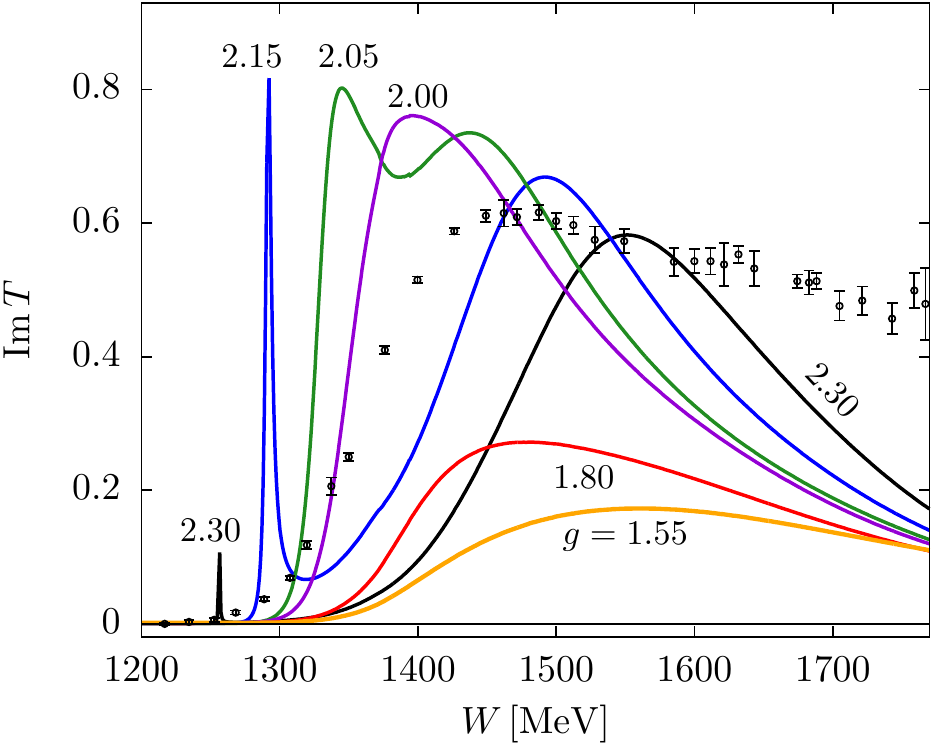}
\end{center}
\vspace{-12pt}
\caption{The imaginary part of the $T$ matrix for $g$ from 1.55 to 2.3
and $\mu_\sigma=\Gamma_\sigma=600$~MeV.}
\label{fig:ImTnoR}
\end{figure}

The $T$ matrix is obtained by solving the Heitler equation 
$T=K + \mathrm{i}KT$.
The influence of the dynamically generated state  on the $T$ matrix is 
best visualized by observing the behaviour of the imaginary part of $T$ 
(Fig.~\ref{fig:ImTnoR}) as we increase $g$. 
For $g$ below the critical value (but not too small)  there is a single 
bump roughly where $w_{\mathrm{min}}$ reaches its minimum. 
For $g=2.05$ we have two poles in the $K$ matrix and two peaks in Im$T$.
As we increase $g$, the lower peak gets narrower and weaker and 
disappears at the location of the two pion threshold. 
The upper peak moves to higher $W$ and becomes wider.

\begin{table}[h]
\caption{\label{tab:SpoleN}%
$S$-matrix pole position and modulus for $g$ from 1.8 to 2.15;
at $g=2.15$ the width and modulus of the lower resonance can not be
reliably determined. The PDG values are taken from \cite{PDG}.}
\begin{ruledtabular}
\begin{tabular}{ccrrc}
$g$ & Re$W_p$ & $-2{\rm Im}W_p$ & $|r|$ & $\vartheta$\\{}
    & [MeV]   & [MeV]   &      &                   \\
\colrule
PDG  & 1370 & 180 & 46    & $-90^\circ$ \\
\colrule
1.55 & 1407 &  207 & 12.6 & $-101^\circ$ \\
1.80 & 1395 &  148 & 10.5 & $-79^\circ$ \\
1.95 & 1382 &  129 & 17.1 & $-59^\circ$ \\
2.00 & 1375 &  111 & 34.0 & $-44^\circ$ \\
\colrule
2.05 & 1331 &   44 &  7.3 & $-62^\circ$ \\  
     & 1438 &  147 & 18.6 & $-17^\circ$ \\  
\colrule
2.15 & 1291 &      &      &            \\  
     & 1476 &  166 & 30.1 & $-27^\circ$ \\  
\end{tabular}
\end{ruledtabular}
\end{table}

From scattering amplitudes we can use the Laurent-Pietarinen expansion
\cite{L+P2013,L+P2014,L+P2015,L+P2014a} to extract the information 
about the $S$-matrix poles shown in Table~\ref{tab:SpoleN} 
which offers a deeper insight into the mechanism of resonance formation.
Notice that the pole in the $S$ matrix emerges already before
the critical value of $g$ is reached.
This means that it is not necessary that the kernel in the 
Lippmann-Schwinger equation (\ref{eq4chi}) becomes singular in order 
to produce a resonance --- or, equivalently, it is not necessary
that the $K$ matrix possesses a pole in the vicinity of the resonance.
The mass of the $S$-matrix pole, Re$W_p$, is close to the 
mass of the Roper pole extracted from the data while the width 
and the modulus are relatively too small.
Above the critical $g$, where the $K$ matrix acquires two poles on the
real axis, two poles appear also in the complex plane with the upper 
pole gaining strength as it moves toward higher $W$ while the opposite 
is true for the lower one as it moves toward lower $W$.

\begin{figure}[h]
\begin{center}
\includegraphics[width=80mm]{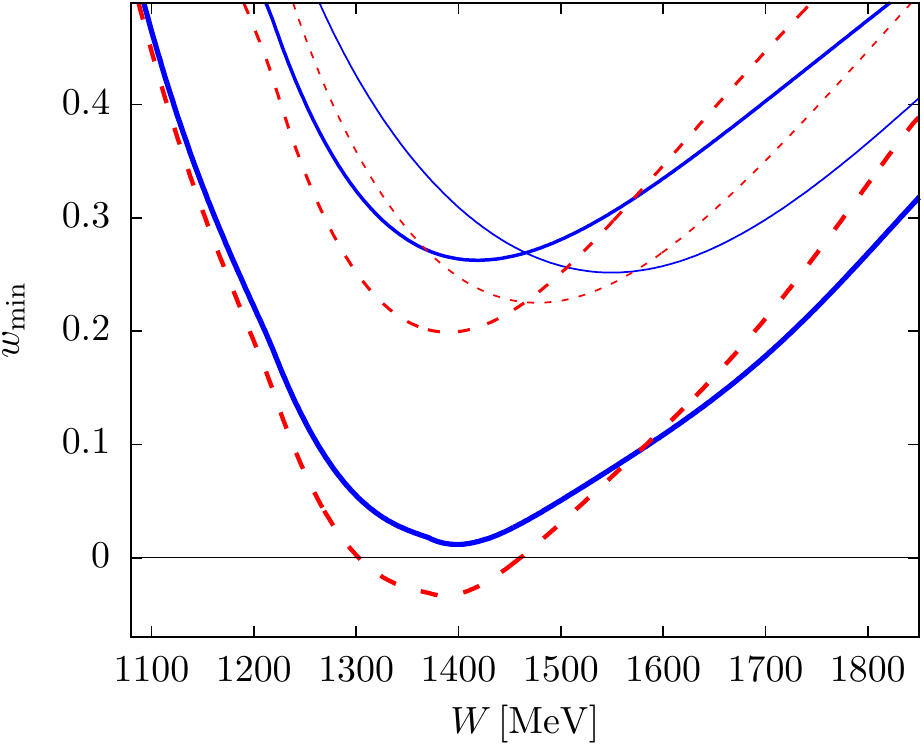}
\end{center}
\vspace{-12pt}
\caption{The lowest eigenvalue of {\sf A} as a function of $W$ for 
$g=2.0$, $\mu_\sigma=\Gamma_\sigma= 600$~MeV (blue) and 
$\mu_\sigma=\Gamma_\sigma=500$~MeV (red); the thick solid lines correspond 
to the full calculation, medium lines to the system without $\Delta$, 
and thin lines to the $\sigma N$ system alone.}
\label{fig:wminNDs}
\end{figure}

Despite its simplicity our model is able to predict correctly
the mass of the Roper resonance.
On the other hand, it is precisely this simplicity that enables us 
to study and reveal the parameters that determine the resonant energy.
At first glance it might seem that it is the  mass of the $\Delta$ 
that most strongly influences the position of the resonance since 
Re$W_p$ almost coincides with the $\pi\Delta$ threshold.
However, increasing/decreasing  the (Breit-Wigner) mass of the $\Delta$
turns out to have very little effect on the position. 
This remains true even if we remove the $\pi\Delta$ channel and 
eliminate completely the $\Delta$ intermediate state from the loops.
The effect can be best observed through the behaviour of the lowest 
eigenvalue of the matrix {\sf A}, $w_{\mathrm{min}}$, as a function of
$W$, for which we have shown that the position of its minimum coincides 
with the mass of the resonance pole even when  $w_{\mathrm{min}}$ does 
not touch or cross zero.
In Fig.~\ref{fig:wminNDs} one sees that the shape of $w_{\mathrm{min}}(W)$
changes little if the $\Delta$ is removed from the calculation: 
in particular the position of the minimum remains almost unchanged.
In fact, by increasing the coupling strength $g$ in the case of 
only two channels, the two curves would almost coincide.
The conclusion is further supported by analysing the parameters of 
the $S$-matrix pole in Table~\ref{tab:SpoleNsi}: for $g=2.00$ the mass 
remains close to the corresponding three-channel case in 
Table~\ref{tab:SpoleN}, while its width is increased and the modulus 
decreased, in agreement with the tendency shown in Table~\ref{tab:SpoleN}
when reducing the coupling strength in the three-channel case.
Similarly, for a larger value of $g=2.25$ the mass and the width 
are reduced and the modulus increased.

\begin{table}[h]
\caption{\label{tab:SpoleNsi}%
Same as table~\ref{tab:SpoleN} for the case without the
$\pi\Delta$ channel.}
\begin{ruledtabular}
\begin{tabular}{ccrrc}
$g$ & Re$W_p$ & $-2{\rm Im}W_p$ & $|r|$ & $\vartheta$\\{}
    & [MeV]   & [MeV]   &      &                    \\
\colrule
PDG  & 1370 & 180 & 46    & $-90^\circ$ \\
\colrule
2.00 & 1342 &  285 & 18.8 & $-11^\circ$ \\
2.25 & 1329 &  204 & 28.5 & $-125^\circ$ \\
\end{tabular}
\end{ruledtabular}
\end{table}

Finally, our model allows us to switch off the $\pi N$ interaction
and study the $\sigma N$ system alone.
In this case the minimum of $w_{\mathrm{min}}(W)$ is shifted higher in
$W$, slightly above the $\sigma N$ threshold
(for the nominal mass of the $\sigma$ meson), see Fig.~\ref{fig:wminNDs}.
Our model therefore proposes the following scenario for 
the formation of the Roper resonance: the $\sigma N$ interaction is
responsible to generate a quasi-bound state close to the  $\sigma N$
threshold; by coupling this state to the $\pi N$ state, the
energy of the quasi-bound state is reduced to around 1400~MeV;
furthermore, the coupling to the $\pi\Delta$ system makes the
system more bound (or, alternatively, produces the quasi-bound
state for weaker couplings) but does not change its position.
While the position of the first state (in the $\sigma N$ channel alone) 
still strongly depends on the (nominal) $\sigma$ mass, the final state 
is only weakly sensitive to the variation of the $\sigma$ mass.

\begin{figure}[h]
$$\includegraphics[width=80mm]{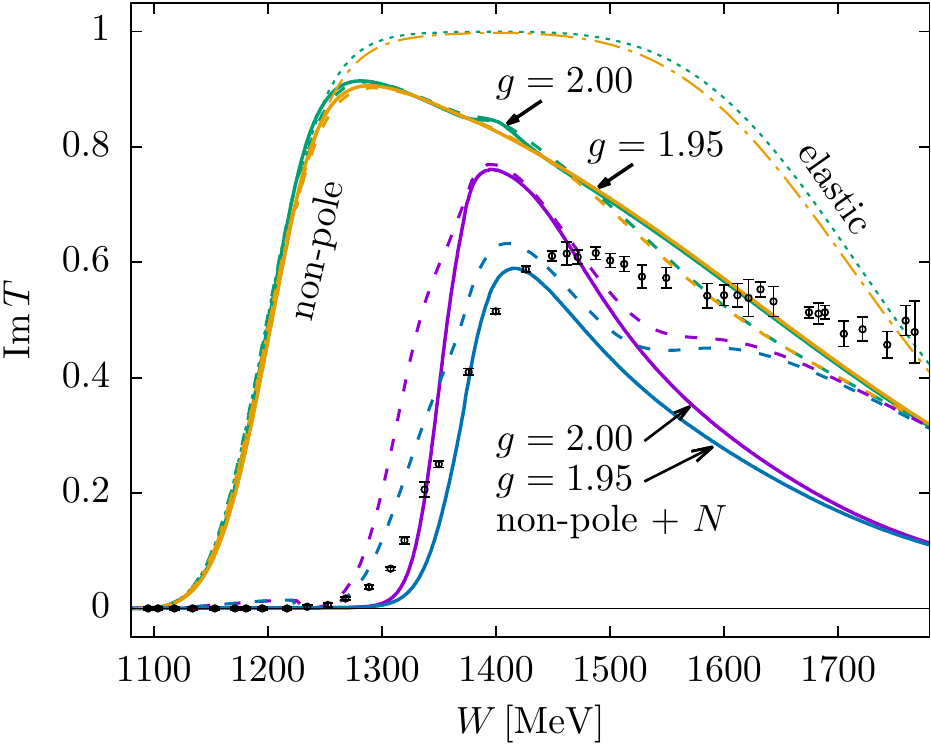}$$
\vspace{-12pt}
\caption{The imaginary part of the $T_{\pi N\,\pi N}$ amplitude;
the non-pole contribution is shown separately as well as 
the contribution from the lowest state of the {\sf A} matrix
(dashed lines) for $g=2.00$ and $g=1.95$.}
\label{fig:ImTnoRN}
\end{figure}

If we remove the nucleon pole and keep only the background term
(see Fig.~\ref{fig:ImTnoRN}) we encounter a similar situation as 
discussed in Ref.~\cite{Roenchen13}.
As mentioned in relation with Eq.~(\ref{KnoR}) both the (positive) 
non-pole contribution to the $K$-matrix and the (negative) 
nucleon-pole contribution are proportional to the inverse of the lowest 
eigenvalue  of {\sf A} which reaches its minimum at around 1400~MeV.
Consequently the $K$-matrix elements of both parts acquire 
very large values which almost cancel in the resulting $K$ matrix.
This cancellation is reflected in a high sensitivity of 
the Im$T_{\pi N\,\pi N}$ to small variations of the model parameters 
as can be seen in Fig.~\ref{fig:ImTnoRN}.
This is not the case with the Im$T_{\pi N\,\pi N}$ calculated from 
the non-pole part alone; 
it rises quickly towards unity (as indicated by the
elastic matrix element alone). 
Above the two-pion threshold, the other two channels start to 
contribute, resulting in a gradual decrease of its value.
The resulting maximum has therefore no physical meaning whatsoever.

Let us mention that the mechanism of the resonance formation carries 
some similarities with the model considered by the Coimbra-Ljubljana 
group \cite{Pedro} using the quark-level Chromodielectric Model which 
incorporates the linear $\sigma$ model with an additional dynamical 
(chromodielectric) field responsible for the quark binding.
The stability condition amounts to solving the Klein-Gordon equation 
for the $\sigma$-meson modes.
The lowest mode turns out to be some 100~MeV below the $\sigma N$ 
threshold, similarly as in the present case.
In that calculation, the $\sigma$ mass was higher 
(i.e. 700~MeV and 1200~MeV) than in our case and the corresponding 
bound state was above the $2s$ quark excitation; consequently, 
the Roper resonance was interpreted as a linear superposition of 
the dominant quark excitation and a quasi-bound $\sigma N$ state.


\section{\label{sec:withR} Including a three-quark resonant state}

We now include in (\ref{PsiH})  a three-quark configuration with one
quark excited to the $2s$ state.
The coupling of this state to $\pi N$ and $\pi \Delta$ is calculated
in the underlying quark model, while the  $\sigma NR$ coupling
is assumed to be equal to the  $\sigma NN$ coupling.

The meson amplitude (proportional to the $K$ matrix) now takes 
the form
\begin{equation}
   \chi_{\alpha\delta}(k,k_\delta) = 
    c_{\delta N}{\cal V}_{\alpha N}(k)+ c_{\delta R}{\cal V}_{\alpha R}(k)
    + {\cal D}_{\alpha\delta}(k,k_\delta)\,,
\end{equation}
with ${\cal V}_{\alpha N}$ and $ {\cal D}_{\alpha\delta}(k,k_\delta)$
satisfying ({\ref{eq4VN}) and  ({\ref{eq4D}), respectively, and
\begin{equation}
\mathcal{V}_{\alpha R}(k)
= {V}_{\alpha R}(k)
 + \sum_{\beta}  \int\mathrm{d}k'\;
        {\mathcal{K}_{\alpha\beta}(k,k')
         \mathcal{V}_{\beta R}(k')
   \over   \omega_\beta(k')+E_\beta(k')-W}\,.
\label{eq4DX}
\end{equation}
The nucleon and the three-quark ($3q$) resonant state mix through meson 
loops, yielding the following set of equations for $c_{\alpha N}$ and
$c_{\alpha R}$:
 \begin{equation}
\begin{array}{lll}
G_{RR}(W)\,c_{\alpha R} & +\; G_{RN}(W)\,c_{\alpha N} & 
         = \mathcal{V}_{\alpha R}(k_\alpha)\,,\\
G_{NR}(W)\,c_{\alpha R} & +\; G_{NN}(W)\,c_{\alpha N} & 
         = \mathcal{V}_{\alpha N}(k_\alpha)\,,
\label{eq4c}
\end{array}
\end{equation}
with $G_{NN}$ given in (\ref{eq4lamN}), and
\begin{eqnarray*} 
G_{RR}(W) &=& W - m_R^0
       + \sum_\beta\int\mathrm{d}k\;{\mathcal{V}_{\beta R}(k){V}_{\beta R}(k)
                \over \omega_\beta(k)+E_\beta(k)-W}\,,
\\
G_{NR}(W) &=& G_{RN}(W)=
\sum_\beta\int\mathrm{d}k\;{\mathcal{V}_{\beta N}(k){V}_{\beta R}(k)
                \over \omega_\beta(k)+E_\beta(k)-W}\,, 
\end{eqnarray*}
were $m_R^0$ is the bare mass of the $3q$ resonant state.

Let ${\sf U}$ denote the unitary transformation that diagonalises 
the ${\sf G}$ matrix in the left-hand side of (\ref{eq4c}):
\begin{equation}
    {\sf U}{\sf G}{\sf U}^T = \hbox{diag}[\lambda_R(W), \lambda_N(W)]\,.
\label{eq4U}
\end{equation}
The resonance part of the $K$ matrix can then be cast in the form
\begin{equation}
   K_{\alpha\beta}^{\mathrm{res}} 
  = \pi\,\mathcal{N}_\alpha\mathcal{N}_\beta\left[
    {\widehat{\cal V}_{\alpha R}\widehat{\cal V}_{\beta R}\over \lambda_R(W)}
  + {\widehat{\cal V}_{\alpha N}\widehat{\cal V}_{\beta N}\over \lambda_N(W)} 
    \right]\,,
\end{equation}
where
\begin{equation}
    \widehat{\cal V}_{\alpha R} = 
         u_{RR}{\cal V}_{\alpha R} + u_{RN}{\cal V}_{\alpha N}\,,
\quad 
    \widehat{\cal V}_{\beta N} = 
         u_{NR}{\cal V}_{\beta R} + u_{NN}{\cal V}_{\beta N}
\label{calVmix}\,,
\end{equation}
and $u$'s denote the matrix elements of ${\sf U}$.
The $3q$ resonant state contains an admixture of the 
ground state and vice versa.
Due to the particular ansatz for the meson amplitude (\ref{barVN}), 
$\widehat{\cal V}_{\beta N}$ vanishes at the nucleon pole ($W=m_N$).
Also, due to this ansatz,  one of the zeros of $\lambda_N(W)$
is always at the nucleon mass, $m_N$.

The set of equations (\ref{eq4x}) and (\ref{eq4z}) is supplemented by 
an equation for $\mathcal{V}_{\alpha R}(k)$ which has the same form as 
(\ref{eq4x}) with  $V_{\beta R}$ replacing $\bar{V}_{\beta N}$  on the right.
It is important to notice that the matrices {\sf M}, given by (\ref{eq4M}), 
and ${\sf A}={\sf I} +  {\sf M}$ remain unchanged with respect to the case 
with no $3q$ resonant state and hence for $g$ larger than the critical $g$, 
two poles in the $K$ matrix appear at the same energies as in the case 
without the $3q$ resonant state; other poles appear at the zeros of 
$\lambda_N$ and $\lambda_R$.

The free parameter of our model is the bare mass of the $3q$ 
resonant state, $m_R^0$. 
In our calculation we do not fix it but adjust it in such a way 
that one of the zeros of $\lambda_R(W)$ (poles of the $K$ matrix) 
is kept at the prescribed value $m_R$.
In our analysis we therefore study the influence of $m_R$ on the
behaviour of the scattering amplitudes.
A similar model of the P11 scattering has been studied in 
\cite{EPJ2008,EPJ2009} as well as in \cite{EPJ2016}.
In these calculations we kept $m_R$ close to the Breit-Wigner mass
and included the $\sigma N$ channel only at the tree level
(ignoring the $\sigma$-meson loops).

\begin{figure}[h]
\begin{center}
\includegraphics[width=80mm]{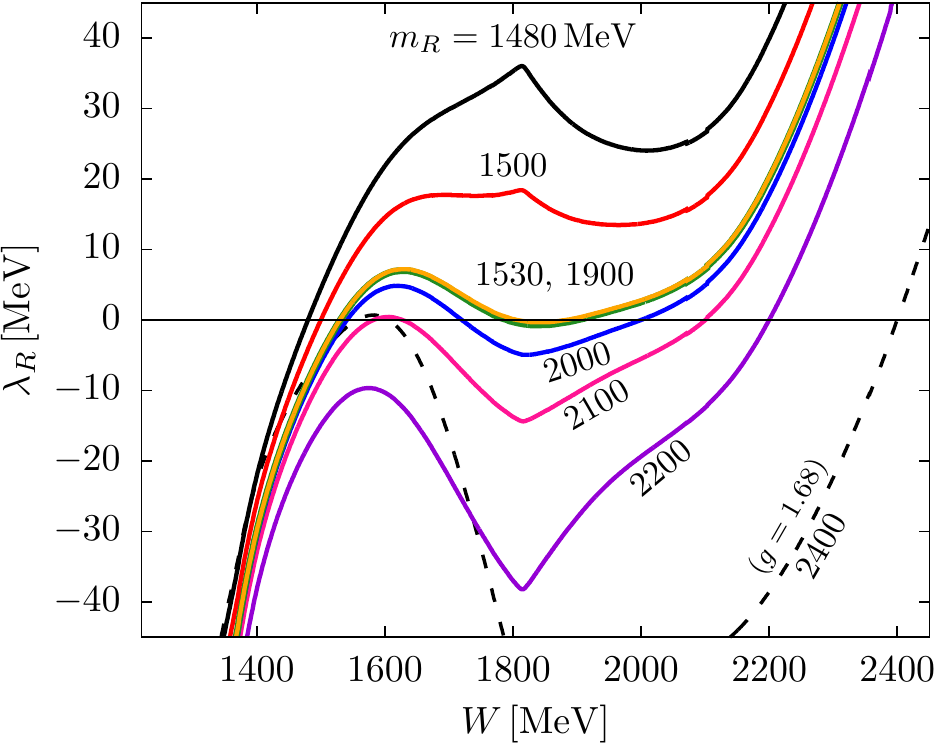}
\end{center}
\vspace{-12pt}
\caption{$\lambda_R(W)$ for masses of the $3q$ resonant state $m_R$ 
from 1480~MeV to 2200~MeV and $g=1.55$, $\mu_\sigma=\Gamma_\sigma=600$~MeV.}
\label{fig:lamR}
\end{figure}

For small $g$, the poles of the $K$ matrix are only at $m_N$ and $m_R$, 
while for larger $g$, $\lambda_R(W)$ may develop additional zeros. 
A typical behaviour of $\lambda_R(W)$ for different $m_R$ is 
displayed in Fig.~\ref{fig:lamR}.
Below $m_R\approx 1520$~MeV, the zero crossing at $m_R$ yields
the only pole in the $K$ matrix; above this value, two additional 
poles appear, for smaller $m_R$ at $W$ higher than $m_R$, while for 
$m_R>1600$~MeV at least one of the additional $K$-matrix poles
emerges below $m_R$.
It is important to stress that these two additional poles
are not directly related to the zeros of $w_{\mathrm{min}}$ discussed in 
the previous section since the value of $g$ is smaller than the 
critical value; nonetheless, the emergence of these two poles 
{\em is\/} a consequence of the dynamical state whose effect 
is enhanced by the presence of the $3q$ resonant state.
The most striking observation is that $\lambda_R(W)$ has very
similar behaviour for $m_R=1530$~MeV as for $m_R=1900$~MeV even
though the origin of the lowest $K$-matrix pole is different.
The resulting poles in the $S$ matrix are displayed in 
Table~\ref{tab:SpoleR}.
The position and the residue of the Roper pole are well reproduced for
$g=1.55$ and $\mu_\sigma=\Gamma_\sigma=600$~MeV for a wide range of 
values of $m_R$ between 1520~MeV and 2000~MeV, and remain close to 
the PDG values even if we considerably alter the values of $g$.
The results are also rather insensitive to a simultaneous reduction 
of the $\sigma$ mass and $g$.
\begin{table}[h]
\caption{\label{tab:SpoleR}%
$S$-matrix pole positions for various values of $g$,
$m_R$ and $\mu_\sigma$ ($\Gamma_\sigma=\mu_\sigma$).}
\begin{ruledtabular}
\begin{tabular}{ccccccc}
$ m_R$ & $\mu_\sigma$ & $g$ & Re$W_p$ & $-2{\rm Im}W_p$ & $|r|$ 
            & $\vartheta$\\{}
  [MeV] &   [MeV]   &      & [MeV]   & [MeV]           &     
            & \\
\colrule
\multicolumn{3}{c}{PDG} & 1370 & 180 & 46  & $-90^\circ$ \\
\colrule
2000  & 600 & 1.55 & 1368 &  180 & 48.0 & $-87^\circ$ \\
2000  & 600 & 1.70 & 1361 &  156 & 41.9 & $-77^\circ$ \\
\colrule
1530  & 600 & 1.55 & 1367 &  180 & 47.5 & $-86^\circ$ \\
2400  & 600 & 1.68 & 1370 &  177 & 42.6 & $-87^\circ$ \\
3000  & 600 & 1.85 & 1364 &  188 & 37.7 & $-98^\circ$ \\
\colrule
2000  & 500 & 1.43 & 1369 &  172 & 40.2 & $-82^\circ$ \\
1530  & 500 & 1.36 & 1365 &  174 & 43.6 & $-82^\circ$ \\
\end{tabular}
\end{ruledtabular}
\end{table}

The fact that the position of the pole remains so stable even if 
we considerably change the parameters of the model clearly shows 
that the position is determined by the  dynamical state discussed 
in the previous section rather than the value of $m_R$.
It almost coincides with the minimum of the lowest eigenvalue 
$w_{\mathrm{min}}$ of the matrix {\sf A}.
Here we encounter a similar situation as in the previous section, 
namely, that the $S$-matrix pole appears where $w_{\mathrm{min}}$ 
(or, equivalently, $\lambda_R$) only approaches zero, i.e. without 
producing poles in the $K$ matrix.
However, the dynamical state alone yields too small values for the 
width and the residuum of the pole; these observables are  brought 
closer to the values reported by PDG by inclusion of a $3q$
resonant state.
The interplay of these two states is displayed in Fig.~\ref{fig:onoff} 
showing the behaviour of Im$T(W)$ for typical values of $g$.
For intermediate values of $g$ which best reproduce the properties of 
the resonance when the $3q$ resonant state in turned on, the effect 
of the dynamically generated state is still weak; for larger values  
of $g$ this state dominates and the influence of the $3q$ resonant 
state is almost invisible.

\begin{figure}[h]
$$\includegraphics[width=80mm]{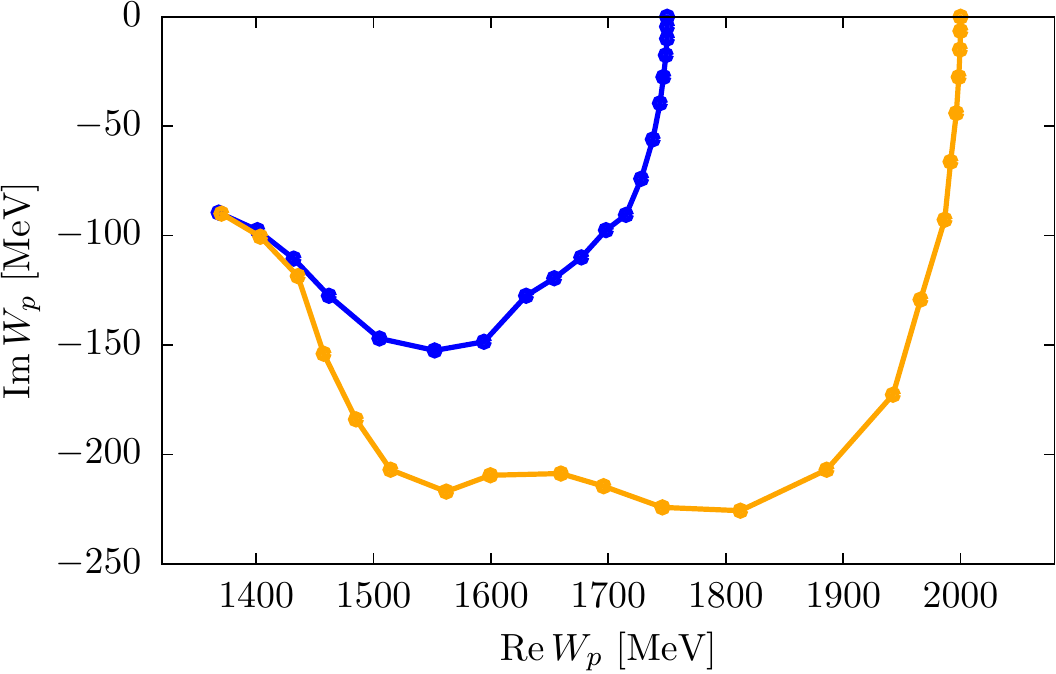}$$
\vspace{-12pt}
\caption{Evolution of the Roper pole as the interaction
strength is gradually switched on for two bare masses of
the three-quark configuration, 1750~MeV and 2000~MeV.}
\label{fig:poles}
\end{figure}
This conclusion is further strengthened by observing the
evolution of the Roper resonance pole as the interaction strength of 
the $\sigma$ as well as the pion interactions  is gradually switched 
on (see Fig.~\ref{fig:poles}), similarly as in Ref.~\cite{Sato-Lee10a}.
In our approach we fix the position of the $K$-matrix pole, 
which in the limit of zero coupling coincides with the energy 
of the bare three-quark state.
We choose two different values, 1750~MeV and 2000~MeV, which are 
in  agreement with the prediction of the continuum approach to 
the baryon bound-state, e.g. as in Ref.~\cite{Segovia15}.
As in the work of the EBAC group the continuous trajectory from
the bare state shows that the resonance indeed originates from
the bare state; nonetheless, the fact that the trajectories from 
two different bare states meet almost at the same point at the value
of $W$ where the dynamically generated ``molecular'' state 
attains its lowest value,
confirms the notion that both states (mechanisms) contribute to 
the formation of the Roper resonance.
This is true even if the $\sigma$ coupling is substantially
weaker compared to the situation treated in the previous section.

The mixing of the ground state to the $3q$ resonant state through the 
meson loops is measured by the squared matrix element $u^2_{RN}$ 
of the ${\sf U}$ matrix introduced in Eq.~(\ref{eq4U}).
The value of  $u^2_{RN}$ strongly depends on $W$ and reaches its maximum 
near the mass of the resonance pole. 
For the typical value of $g=1.55$ it comes close to 50~\% which means 
that the probability of finding the excited three-quark configuration 
in the $3q$ resonant state is further reduced with respect to the 
pure quark model.\footnote{%
Similarly as in the calculations on the lattice, one should be aware 
that it is not possible to directly compare the probabilities for a 
(quasi)-bound three-quark configuration and the meson configurations 
since the latter are proportional to the (infinite) volume.}

\begin{figure*}[t]
\begin{center}
\includegraphics[width=80mm]{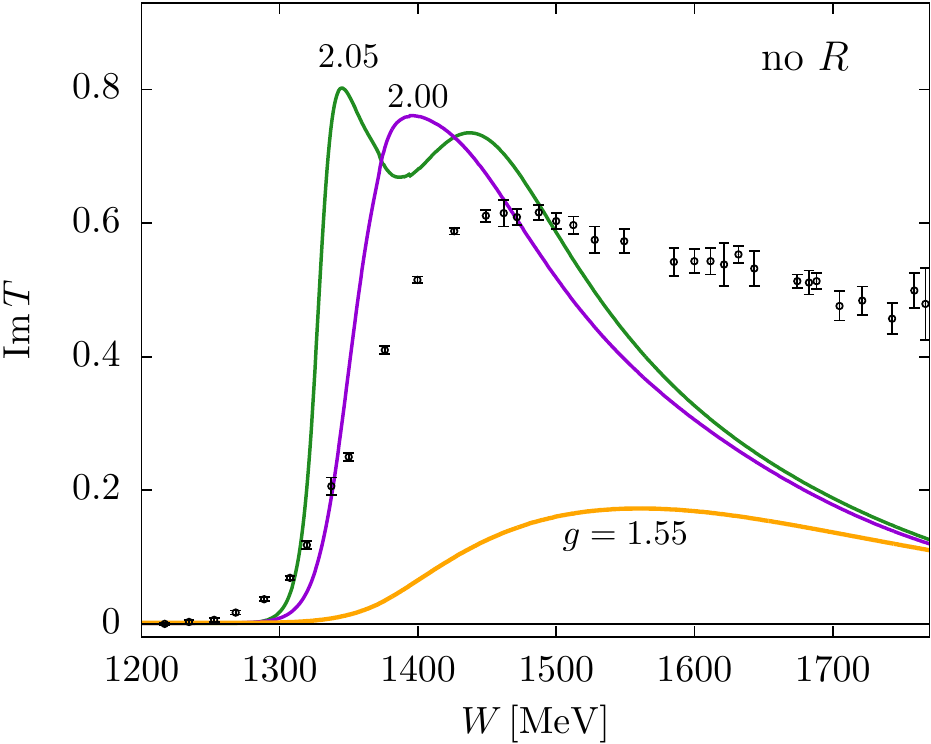}~~~%
\includegraphics[width=80mm]{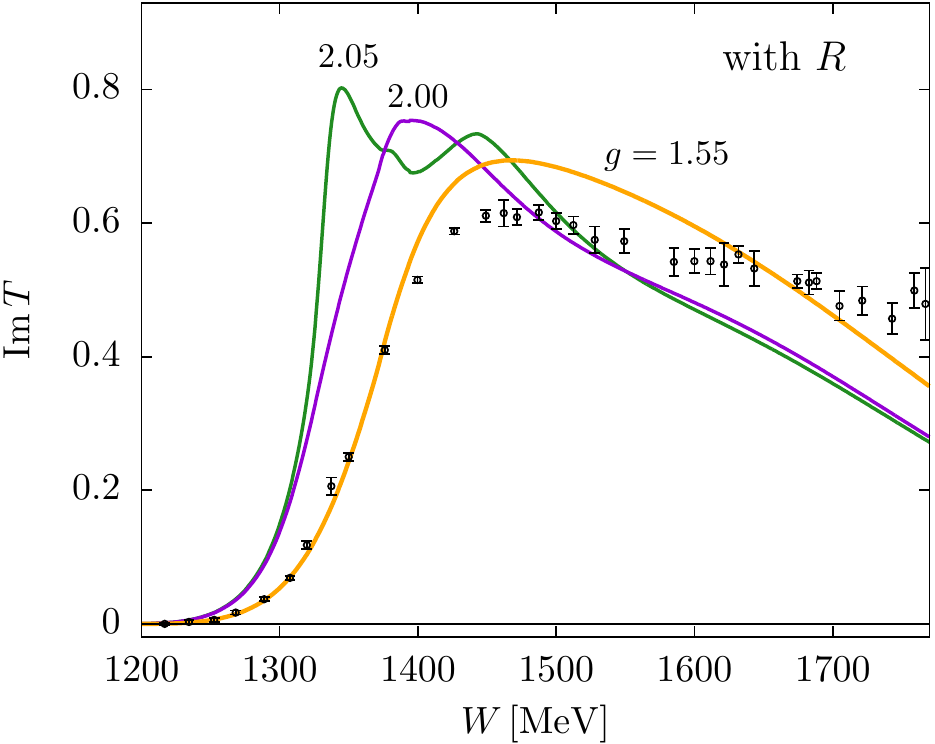}
\end{center}
\vspace{-12pt}
\caption{Im$T(W)$ for $\mu_\sigma=600$~MeV when the $3q$ resonant 
state with $m_R=2000$~MeV is turned off (left) and on (right).}
\label{fig:onoff}
\end{figure*}

Our coupled channel approach is similar to the pioneering approach
of Krehl et al. \cite{Krehl} using a coupled-channel meson
exchange model and that of the Adelaide group using Hamiltonian 
effective field theory.
They both solve the Lippmann-Schwinger equation for the $T$ matrix
which has an analogous form as our $K$ matrix consisting of the
resonant part and the background part.
Krehl et al. \cite{Krehl} were the first to notice that the 
resonance can be formed by using only the $\pi N$ and
the $\sigma N$ channels, which has been confirmed also in our 
calculation in the previous section.
Our results are consistent with those obtained by the Adelaide group.
The advantage of our approach is that it uses a smaller number of 
free parameters since it relates the pion couplings to different 
baryons through the underlying quark model and the calculations in 
other partial waves.
Furthermore, we treat two (or more) baryon states simultaneously and 
thus studying the interplay of the nucleon and the Roper degrees 
of freedom.
The relevance of the $N+\sigma$ admixture to the excited $3q$
configuration has been also realized in the calculation of 
electro-production of the Roper resonance in 
Refs.~\cite{tubingen11,tubingen14}.

\section{\label{conclusion} Conclusion}
We have developed a simplified model including the $\pi N$, $\pi\Delta$ 
and $\sigma N$ channels to study the formation of the Roper resonance 
in the presence of a three-quark resonant state or in its absence.
The number of model parameters is kept as small as possible and only 
the $u$-channel exchange as the sole background process is considered.
The Laurent-Pietarinen expansion has been used to extract the 
$S$-matrix resonance-pole parameters.
Despite the simplicity of the model, the properties of the Roper 
resonance are well reproduced in the intermediate coupling regime in 
which both the dynamically generated state as well as the three-quark 
resonant state contribute.

We have been able to pin down one particular state with $\pi N$, 
$\pi\Delta$ and $\sigma N$ components which dominates the scattering
amplitudes between $W\approx 1300$~MeV  and  $W\approx 1500$~MeV and 
is responsible for the dynamical generation of the resonance.
Its mass lies very close to the mass of the Roper pole
and is very insensitive to rather large variations of the model 
parameters and even to the removal of the $\pi\Delta$ channel.
We infer that this very state determines the mass of the
Roper resonance, except in the case when the three-quark resonant  
state is included with the mass equal to or below 1500~MeV.
Nonetheless, it appears that the dynamically generated state does not 
describe adequately the resonance properties; in the intermediate 
coupling regime it produces only a weak $S$-matrix pole in the complex 
plane which evolves towards the PDG values for the position and the 
residuum only upon inclusion of the three-quark state.
This evolution is rather insensitive to the mass of the three-quark
state which may be as large as 2000~MeV.
In view of the difficulties in the quark model to explain the ordering
of single-particle states in which the $2s$ state would lie lower than 
the $1p$ state, as well as the recent results of the lattice calculations 
\cite{lang16,kiratidis17} which have not found a sizable three-quark 
component  below 1.65~GeV and 2.0~GeV, respectively, the presented model 
appears to rule out the existence of a three-quark resonant state around 
or below 1500~MeV. 
It favours the picture in which the mass of the $S$-matrix pole is 
determined by the energy of the dynamically generated state while its 
width and modulus are strongly influenced by the three-quark resonant 
state.

Though the description of the Roper resonance as a purely dynamically 
generated phenomenon could be further refined by including a richer
set of backgrounds, we can not find a convincing reason to a priori 
exclude its genuine three-quark component.  From our viewpoint
this appears to be the simplest addition needed to reach satisfactory 
agreement with resonance properties extracted from the data.

\begin{acknowledgments}
The Authors wish to thank
Sa\v{s}a Pre\-lov\-\v{s}ek for stimulating and fruitful
discussions.
\end{acknowledgments}

\appendix*
\section{\label{appA} Structure of the {\sf A} matrix}

The {\sf A} matrix elements are of the form $A^{\beta\eta}_{\alpha i,\gamma j}$,
and the vectors are labeled by the three indices $\beta\alpha i$.
Below $N, \Delta$ and $\sigma$ are a shorthand notation for the 
$\pi N, \pi\Delta$ and $\sigma N$ channels, while $i$ and $j$ denote 
the intermediate baryon, $N$ or $\Delta$ (see Fig.~\ref{fig:Agraph}).
The index $i$ ($j$) is dropped in those matrix elements that
involve the $\sigma N$ channel since the $\sigma$ vertex preserves
spin/isospin and only $N$ or $\Delta$ is present.
The dimension of the first two submatrices is 5, while that of
the third one is 3 resulting in $\mathrm{dim}({\sf A})=13$.

\newcommand{\bstrut}{\vrule width 0pt height 12pt depth 5pt}
\begin{widetext}
$$
\setlength{\tabcolsep}{9pt}        
\begin{array}{|c|ccc|ccc|ccc|}
\hline
\lower1pt\hbox{${\beta\alpha i}$}\hbox{\LARGE$\diagdown$}\kern-3pt
\raise4pt\hbox{${\eta\gamma j}$}
& NNj
& N\Delta j
& N\sigma N
& \Delta Nj
& \Delta \Delta j
& \Delta \sigma \Delta 
& \sigma NN
& \sigma \Delta \Delta 
& \sigma \sigma N
\\
\hline
\bstrut
NNi 
& \delta_{i,j}+M^{N}_{Ni,Nj}
& M^{N}_{Ni,\Delta j}
& M^{N}_{Ni,\sigma }
& 0
& 0
& 0
& 0
& 0
& 0
\\
\bstrut
N\Delta i
& M^{N}_{\Delta i,Nj}
& M^{N}_{\Delta i,\Delta j}
& M^{N}_{\Delta i,\sigma}
& \delta_{i,j}
& 0
& 0
& 0
& 0
& 0
\\
\bstrut
N\sigma N
& M^{N}_{\sigma,Nj}
& M^{N}_{\sigma,\Delta j}
& M^{N}_{\sigma,\sigma}
& 0
& 0
& 0
& 1
& 0
& 0
\\
\hline
\bstrut
\Delta Ni
& 0
& \delta_{i,j}
& 0
& M^{\Delta}_{Ni,Nj}
& M^{\Delta}_{Ni,\Delta j}
& M^{\Delta}_{Ni,\sigma }
& 0
& 0
& 0
\\
\bstrut
\Delta \Delta i
& 0
& 0
& 0
& M^{\Delta}_{ \Delta i,Nj}
& \delta_{i,j}+M^{\Delta}_{\Delta i,\Delta j}
& M^{\Delta}_{ \Delta i,\sigma }
& 0
& 0
& 0
\\
\bstrut
 \Delta\sigma \Delta 
& 0
& 0
& 0
& M^{\Delta}_{\sigma, Nj}
& M^{\Delta}_{\sigma, \Delta j}
& M^{\Delta}_{\sigma, \sigma}
& 0
& 1
& 0
\\
\hline
\bstrut
\sigma NN
& 0
& 0
& 1
& 0
& 0
& 0
& M^{\sigma}_{N,N}
& M^{\sigma}_{N,\Delta}
& M^{\sigma}_{N,\sigma}
\\
\bstrut
\sigma\Delta\Delta 
& 0
& 0
& 0
& 0
& 0
& 1
& M^{\sigma}_{\Delta, N}
& M^{\sigma}_{\Delta, \Delta}
& M^{\sigma}_{\Delta, \sigma}
\\
\bstrut
\sigma \sigma N
& 0
& 0
& 0
& 0
& 0
& 0
& M^{\sigma}_{\sigma, N}
& M^{\sigma}_{\sigma, \Delta}
& 1+M^{\sigma}_{\sigma, \sigma}
\\
\hline
\end{array}
$$
\end{widetext}

\end{document}